# INHOMOGENEOUS BROADENING OF EXCITON LINES IN MAGNETO-OPTICAL REFLECTION FROM CdTe/CdMgTe QUANTUM WELLS

G. V. Astakhov, V. A. Kosobukin, V. P. Kochereshko,

*A.F.Ioffe Physical-Technical Institute, Russian Academy of Sciences, 194021 St.Petersburg, Russia*

D. R. Yakovlev*, W. Ossau, G. Landwehr

*Physikalisches Institut der Universität Würzburg, 97074 Würzburg, Germany*

T. Wojtowicz, G. Karczewski, J. Kossut,

*Institute of Physics, Polish Academy of Sciences, PL02668 Warsaw, Poland*

**Abstract.** An approach is proposed to determine parameters of homogeneous and inhomogeneous broadening for reflection lines of quasi-two-dimensional excitons in semiconductor quantum well structures. A phenomenological model is developed taking into account fluctuations of the exciton energy which are expected to cause inhomogeneous broadening of exciton reflectivity lines. This concept is applied to magneto-optical studies of CdTe/(Cd,Mg)Te single quantum well heterostructures containing a two-dimensional electron gas. By the analysis of excitonic reflection spectra in magnetic fields the parameters of homogeneous and inhomogeneous broadening of the exciton as well as their temperature dependencies were obtained.

# 1. Introduction

Optical reflection and transmission spectroscopy is widely used in the investigation and in the characterization of bulk semiconductors and semiconductor heterostructures. These methods are powerful tools to determine refractive and absorption indices which are the basic characteristics the interaction of light with solids [1]. In particular, resonant reflection (transmission) spectra provide rich information about transition energies, homogeneous broadening and radiative decay rates of excitons as well as the influence of applied electrical and magnetic fields [3-4].

If the homogeneous and inhomogeneous spectral line widths are of the same order, an analysis of the reflectance spectra becomes very complicated and the parameters involved are uncertain.

In order to discriminate between homogeneous and inhomogeneous contributions we suggest to vary the experimental conditions in such a way, that only one of the mechanisms is affected, which will allow to separate both contributions. It has been shown in the past [5] that variation of the temperature allows to determine homogeneous and inhomogeneous broadening independently. However, in semiconductors a temperature increase may lead to the ionization of shallow impurities and may result in additional inhomogeneous broadening due to a modification of electric field fluctuations. Also, an applied magnetic field alone does not allow to achieve our purpose because it could lead to exciton-localization and affect the homogeneous and inhomogeneous linewidth simultaneously [4].

The present approach is based on the observation that the width of the exciton reflection (transmission) line in QW structures containing a 2D electron gas in magnetic fields is different for two opposite circular polarizations of the probing light [6]. This effect has been explained in Refs.[7-9] by spin-dependent exciton-electron scattering.

It is reasonable to assume that inhomogeneous broadening of the exciton linewidth does not depend on the polarization of the probing light and, that consequently the polarization effect gives information about the homogeneous linewidth.

By measuring the exciton reflectivity spectra in two circular polarizations one can extract separately contributions to the total linewidth of polarization-dependent and polarization-independent mechanisms of line broadening. Therefore, it is possible to determine independently the homogeneous and inhomogeneous components of the linewidth.

## 2. Theoretical background

We consider a heterostructure with a single quantum well (SQW) located at a distance $L$ from the surface of a sample as is shown by the inset in Fig.1a. Usually, in interpreting reflectivity spectra of a QW its interfaces are assumed to be perfectly flat [2,3,10]. To describe the reflectivity in the vicinity of an exciton resonance the expression

$$R(\omega - \omega_0) = R_b \left\{ 1 - \frac{8n}{n^2 - 1} \text{Re}[r \cdot \exp(i\Phi)] \right\} \quad (1)$$

is conventionally used, see p.180 in Ref.[3]. In Eq.(1) $R_b = [(n-1)/(n+1)]^2$ is the reflection coefficient of a semiconductor sample with the background refraction index $n$ and without a SQW, $\Phi = 2n\omega L/c$ stands for a phase, and multiple reflections of light from the SQW are neglected. The term

$$r = \frac{i\Gamma_0}{\omega_0 - \omega - i(\gamma + \Gamma_0)}, \quad (2)$$

depending on the incident light frequency $\omega$, is, in essence, the amplitude reflection coefficient of the SQW put in unbounded semiconductor background, $\omega_0$ is the resonance frequency, $\Gamma_0$ is the radiative and $\gamma$ nonradiative decay rate of a quasi-2D exciton of the SQW.

We generalize the formulas (1,2) by taking into account multiple reflections of light from a SQW and the nearby surface of the sample. Applying the results of [10] to such SQW one can express the power reflection coefficient at normal light incidence as follows:

$$R(\omega - \omega_0) = R_b \left\{ 1 + \frac{\Gamma_0 [A(\omega_0 - \omega) + B\gamma + C\Gamma_0]}{(\omega_0 + \delta\omega_0 - \omega)^2 + (\gamma + \Gamma_0 + \delta\Gamma_0)^2} \right\}. \quad (3)$$

The resonant term in Eq.(3) is associated with a quasi-2D exciton. The corrections

$$\delta\omega_0 = \Gamma_0 \frac{n-1}{n+1} \sin(2n\omega L/c), \qquad \delta\Gamma_0 = \Gamma_0 \frac{n-1}{n+1} \cos(2n\omega L/c) \quad (4)$$

occur due to the presence of the sample surface at a distance $L$ from the middle-plane of SQW. The other parameters in Eq.(3) are expressed as follows

$$\{A, \ B, \ C\} = \frac{8n}{n^2 - 1} \left\{ \sin\Phi, \ \cos\Phi, \ \frac{n^2 + 1}{n^2 - 1} + \cos\Phi \right\}, \quad (5)$$

and the above phase $\Phi = 2n\omega L/c$ entering the parameters of Eqs.(4) and (5) originates from nothing but the presence of the sample surface. Typical reflection spectra calculated along Eq.(3) at different SQW/surface separations $L$ are presented in Fig.1a, which shows a significant variation of the lineshapes with $L$. In turn, Fig.1b illustrates the variations $\delta\omega_0$ and $\delta\Gamma_0$ calculated from Eq.(4) as functions of $L$. The curves in both Figs.1a and 1b depend periodically on $L$ through the phase $\Phi$ due to the interference that occurs between the direct wave and the backward wave reflected from the SQW. Note that the oscillation effects presented in Fig.1b are similar to those discussed earlier in [11].

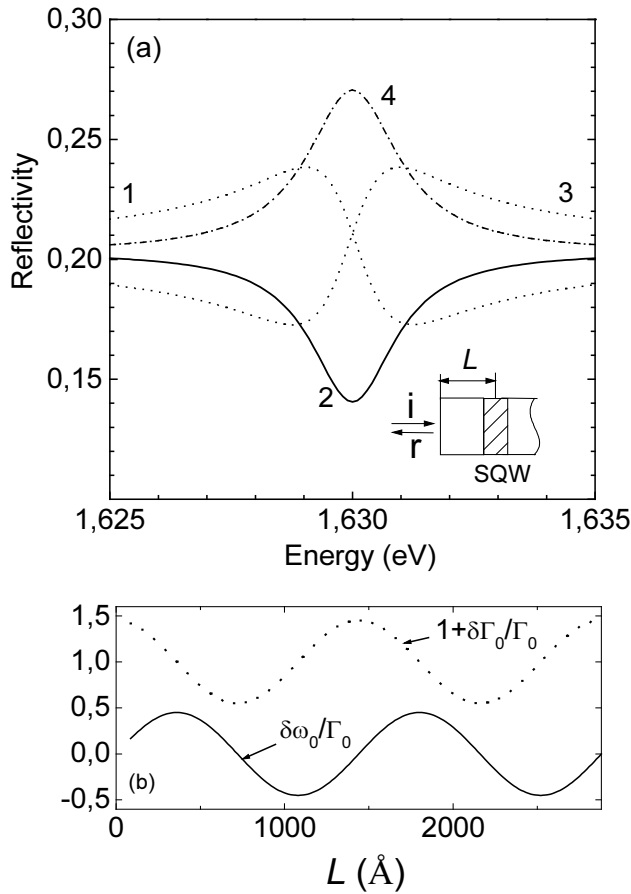

**Fig. 1.** (a) Reflection spectra calculated from Eq.(3) for a SQW with flat interfaces at the following separations $L$ between SQW and sample surface: 360 Å ($\Phi=\pi/2$) - (1), 720 Å ($\Phi=\pi$) - (2), 1080 Å ($\Phi=3\pi/2$) - (3), 1440 Å ($\Phi=2\pi$) - (4). Inset: scheme of a heterostructure with a SQW, incident (i) and reflected (r) light waves being shown. (b) Normalized surface-induced corrections $\delta\omega_0/\Gamma_0$ (solid) and $1+\delta\Gamma_0/\Gamma_0$ (dashed) calculated from Eq.(4) as functions of $L$. Used parameters $n=2.64$, $\hbar\omega_0=1.63$ eV, $\hbar\gamma=1$ meV, $\hbar\Gamma_0=0.1$ meV are typical for a CdTe/Cd$_{0.7}$Mg$_{0.3}$Te SQW with the thickness $a=80$ Å.

Only the homogeneous broadening due to both dissipative and radiative decays of excitonic state is dealt with above, the corresponding linewidths being related to a finite exciton lifetime through the uncertainty principle [12].

Inhomogeneous broadening of an optical excitation is generally ascribed to spatial fluctuations of its resonant energy over a sample [12-14]. In treating the inhomogeneous broadening effects the statistical independence of optical spectra taken from different areas of a disordered sample is of principal importance, because the total optical response is the sum of corresponding intensities [12,13]. In real QW structures homogeneously broadened excitonic spectral lines are known to broaden inhomogeneously due to unavoidable imperfections of the structure [15,16]. An example is the random roughness of QW interfaces leading to QW thickness fluctuations [15-21], the scale of which is comparable with the thickness of an atomic monolayer.

Thus one may consider a QW as an array of areas (islands), each having nearly constant thickness which fluctuates, however, in passing from one island to another. In islands whose lateral sizes are comparable with or larger than the Bohr radius of a quasi-2D exciton the latter may be considered as identical with that typical of an infinite QW having the same thickness (compare with [15,16,21]). Then, fluctuations of the QW thickness affect the energy of an exciton resonance mainly through the confinement effect. Inasmuch as a great number of random islands is probed simultaneously in an optical experiment with a QW, those should be treated as a statistical ensemble whose elements possess random exciton frequencies. Then, an averaged reflection coefficient of a sample

$$\overline{R}(\omega - \overline{\omega}_0) = \int_{-\infty}^{\infty} dv \cdot R(\omega - v) \cdot f(v - \overline{\omega}_0) \quad (6)$$

is obtained on convoluting Eq.(3) with a distribution function $f(v - \overline{\omega}_0)$ of the excitonic transition frequency $\omega_0$ normalized as $\int_{-\infty}^{\infty} dv \cdot f(v - \overline{\omega}_0) = 1$.

The statistical properties of both interfaces and the related exciton transition frequencies are unknown for a QW. Therefore we use Gaussian statistics, i.e.

$$f(v - \overline{\omega}_0) = \frac{1}{\sqrt{2\pi}\Delta} \exp\left(-\frac{(v - \overline{\omega}_0)^2}{2\Delta^2}\right), \quad (7)$$

with $\Delta$ being referred to as a parameter of inhomogeneous broadening. This kind of statistics seems to be most appropriate when many weak random scatterers which are statistically independent of each other are probed. Using measurements of elastic light scattering via excitonic states with random character, a Gaussian correlation has been verified for rough semiconductor surfaces [22]. After substitution of Eqs. (3) and (7) into Eq.(6) one obtains

$$\overline{R}(\omega - \overline{\omega}_0) = R_b \left\{ 1 + \sqrt{\frac{\pi}{2}} \frac{\Gamma_0}{\Delta} \left[ \frac{-A \cdot \delta\omega_0 + B \cdot \gamma + C \cdot \Gamma_0}{\gamma + \Gamma_0 + \delta\Gamma_0} \cdot \operatorname{Re} w(z) - A \cdot \operatorname{Im} w(z) \right] \right\}. \quad (8)$$

Here,

$$w(z) = (i/\pi) \int_{-\infty}^{\infty} dt \cdot \exp(-t^2) \cdot (z - t)^{-1}$$

is the probability integral of the complex argument [23] that is equal to

$$z = \frac{1}{\sqrt{2}\Delta} \left[\omega - \overline{\omega}_0 + i(\gamma + \Gamma_0 + \delta\Gamma_0)\right]. \quad (9)$$

It follows from Eq.(8) that an inhomogeneously broadened line contour is the simplest at $A = 0$, a condition which is fulfilled in accordance with Eq.(5), if $L = N\pi c/(2n\omega_0)$ with integer $N$. Under this condition Eq.(8) describes the Voigt contour [13], if $\Delta \neq 0$, that transforms into the Lorentzian-based function (3), if $\Delta \to 0$.

## 3. Experiment

As a model system we used 80 Å CdTe/Cd$_{0.7}$Mg$_{0.3}$Te single quantum well structures. The samples were grown by molecular beam epitaxy on (100)-oriented GaAs substrates. These structures were modulation-doped with iodide and have a spacer thickness of 100 Å. The concentration of the two-dimensional electron gas (2DEG) was $8\times 10^{10}$ cm$^{-2}$. The structures were covered by a 750 Å Cd$_{0.7}$Mg$_{0.3}$Te cap layer. In our samples with $n = 2.64$, the cap layer thickness $L$=750 Å was chosen to satisfy the above-mentioned condition $A = 0$ in Eq.(8) which corresponds to curve 2 in Fig.1a or, in other words, to the Voigt reflectivity contour. In this case the reflectance lineshape analysis becomes much simpler. Note that a 2DEG is used here to induce a specific excitonic effect in magnetic fields, however, its own contribution to the optical response is estimated to be negligible.

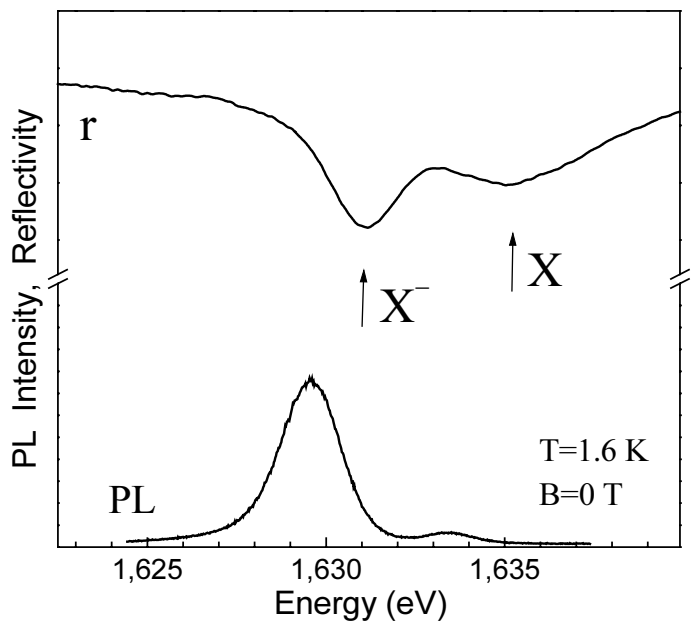

**Fig. 2.** Reflectivity (r) and photoluminesence (PL) spectra of an 80-Å-thick CdTe/Cd$_{0.7}$Mg$_{0.3}$Te SQW measured at $T$=1.6 K.

Figure 2 illustrates a reflection (r) and a photoluminesence (PL) spectrum taken from the sample with a free electron concentration of $8\times10^{10}$ cm$^{-2}$. Both spectra were measured at 1.6 K. The spectra consist of two lines $X$ and $X^-$, separated by 4 meV. The $X$ line corresponds to an exciton in the QW, the $X^-$ line corresponds to a negatively charged exciton (trion) [6]. Both the $X$ and the $X^-$ line are inhomogeneously broadened. There is a Stokes shift between reflectivity and PL spectra which is caused by QW thickness fluctuations.

The exciton reflection spectrum was measured in a magnetic field of 5.5 T applied in Faraday geometry. The results are shown in Fig.3 for two circular polarizations of incident light: $\sigma^+$ and $\sigma^-$, which correspond to different orientations of the exciton spin momentum: +1 and -1. In $\sigma^+$- polarization two lines $X^-$ and $X$ are observed. The $X^-$ line is undetectable in the $\sigma^-$- polarization because the $X^-$ ground state is a singlet and, consequently, in a magnetic field when the 2DEG is completely spin polarized, it can be observed in one polarization only [6].

The $X$ line is observed at energies of 1.6348 eV and 1.6342 eV for $\sigma^+$- and $\sigma^-$ circular polarizations, respectively. A small difference (about 0.6 meV) between these resonant energies is due to the Zeeman splitting of the exciton levels with g-factor $g = |g_e - g_h| = 1.64$ [24]. The important feature of these spectra is that the amplitude and width of the exciton reflectivity line are different for $\sigma^+$ and $\sigma^-$ polarizations.

In the $\sigma^-$- polarization an additional line appears. This line is attributed to combined exciton-cyclotron resonance (ExCR), which corresponds to the photocreation of an exciton with simultaneous excitation of an additional electron to the upper Landau level [25].

## 4. Discussion

The excitonic parameters (see Eq.(8)) can be determined independently and with high accuracy in the absence of inhomogeneous broadening, i.e. for $\Delta \to 0$. The high accuracy of the determined parameters is based on the fact that each of them describes the characteristic features of the exciton reflectivity line independently from the others. For example at A=0 the resonant frequency $\omega_0$ defines the position of the line maximum, the dissipative damping rate $\gamma$ defines the linewidth and the radiative damping constant $\Gamma_0$ defines the line amplitude [2]. As a result, small fluctuations in one of the parameters have only a small influence on the others.

In the presence of inhomogeneous broadening ($\Delta \neq 0$), one can see in Eq.(6) that the amplitude and width of the exciton reflectivity line depends on $\Delta$ and it becomes impossible to determine the parameters $\Delta$, $\gamma$ and $\Gamma_0$ independently.

To distinguish between these parameters we have modulated the exciton line by an applied magnetic field through its effect on the polarization of the exciton. We assume that the inhomogeneous linewidth does not depend on the exciton spin state and that it is the same in both $\sigma^+$ and $\sigma^-$ polarization. It is natural to assume that the radiative damping constant $\Gamma_0$ is also maintained for both circular polarizations. The radiative decay rates $\Gamma_0^+$ and $\Gamma_0^-$ should coincide with each other [26]. Actually, the radiative decay is defined by an overlap of the electron and hole wavefunctions and can not depend on the angular momentum of the exciton (the problems concerning the exciton oscillator strength in modulation doped QWs have been discussed before in Refs.[7-9]). We also assume that the magnetic field of 5.5 T is too weak to have an effect on the exciton radiative damping and on the inhomogeneous broadening. This allows us to determine the values of $\gamma$, $\Gamma_0$ and $\Delta$ by fitting the reflectivity spectra in two polarizations $\sigma^+$ and $\sigma^-$ with the same $\Gamma_0$ and $\Delta$.

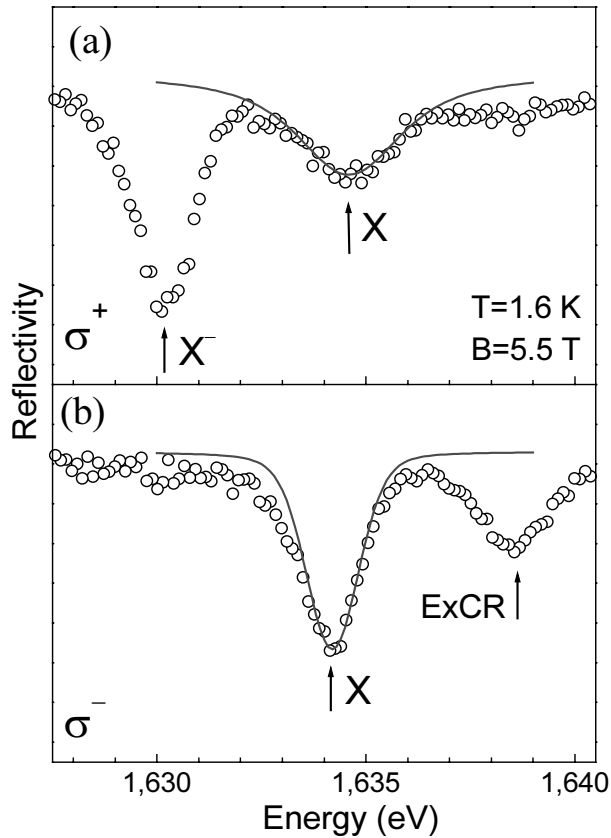

**Fig. 3.** Exciton reflection spectra measured at a temperature of 1.6 K in a magnetic field of 5.5 T for $\sigma^+$ (a) and $\sigma^-$ (b) polarizations. Open circles correspond to experimental data, and solid lines show spectra calculated with Eq.(8).

The above reflection spectra were fitted using Eq.(8) with $\Delta \neq 0$ and assuming $\hbar\Gamma_0^+ = \hbar\Gamma_0^-$. In this case, the best fit values are $\hbar\gamma^+ = 1.1\pm0.08$ meV and $\hbar\gamma^- = 0.15\pm0.08$ meV, while $\hbar\Gamma_0^+ = \hbar\Gamma_0^- = 0.052\pm0.01$ meV were obtained at $\hbar\Delta = 0.56\pm0.08$ meV. The calculated reflection spectra show good agreement with the experimental ones for both $\sigma^+$ and $\sigma^-$ polarizations (Fig.3). Despite of the linewidth dependence on both the parameters $\Delta$ and $\gamma$, the accuracy of the $\Delta$ and $\gamma$ measurements is rather high. This high accuracy arises from the strong polarization dependence of the homogeneous broadening $\gamma$. In undoped structures, in which there is no polarization dependence of the line width, the accuracy of the homogeneous and inhomogeneous linewidth separation is much less.

Assuming the existence of inhomogeneous broadening, we have found the radiative decay rates $\hbar\Gamma_0^+ = \hbar\Gamma_0^- = 0.052$ meV stay constant in the temperature range from 1.6 to 15 K. This finding is very reasonable as radiative damping is determined by an overlap of electron and hole wave functions in excitons, which is not expected to be temperature dependent. In turn this confirms the validity of the used assumptions. The corresponding temperature dependencies of the parameters $\Delta$, $\gamma^+$ and $\gamma^-$ are shown in Fig.4. The inhomogeneous broadening parameter $\Delta$ is found to be nearly independent of temperature. The weak temperature dependence of the $\Delta$ can not be attributed to thermal expansion of the lattice. However, we have to consider that it might be related to an increase of potential fluctuations due to ionization of shallow impurities when the temperature is raised. But this effect is small due to relatively large ionization energies and we suppose that the main reason for the inhomogeneous broadening arises from QW thickness fluctuations.

As far as inhomogeneous broadening is attributed mainly to fluctuations of the QW thickness, one can estimate the r.m.s. roughness height $\xi$ for uncorrelated interfaces through the parameter $\Delta$ as follows

$$\xi = \frac{a}{2\sqrt{2}} \frac{\hbar\Delta}{E_{conf}} \ . \qquad (10)$$

In Eq.(10), $E_{conf} = (\pi\hbar)^2/(2\mu a^2)$ is the carrier confinement energy in the QW with an average thickness $a$, and $\mu$ is the electron-hole reduced mass. Following Ref.[21], relation (10) assumes the parameter $\hbar\Delta$ to be an r.m.s. shift of the exciton transition energy when the confinement energy of the carriers is laterally modulated via the random SQW thickness whose fluctuation is $\sqrt{2}\xi$. Using Eq.(10) with $\hbar\Delta = 0.6$ meV one obtains $\xi \sim 1$ Å which implies that the above observation of inhomogeneous broadening corresponds to a submonolayer fluctuation of the SQW thickness.

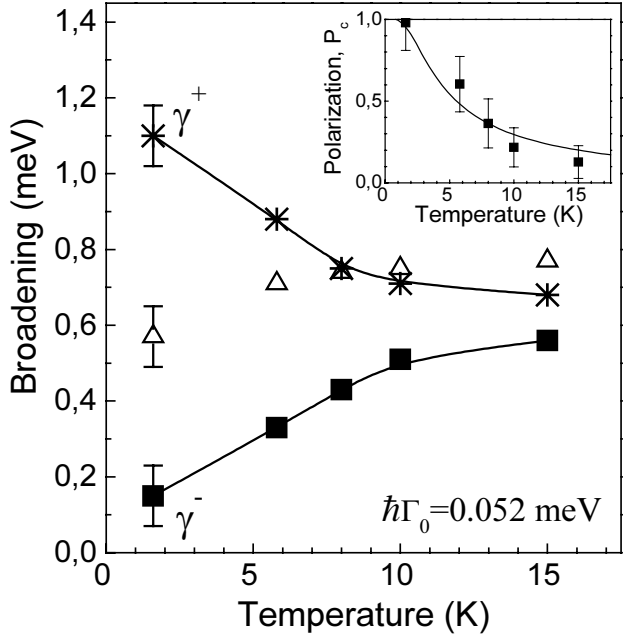

**Fig. 4.** Temperature dependencies of excitonic fitting parameters in a magnetic field of 5.5 T: inhomogeneous broadening $\Delta$ (triangles), and homogeneous broadening $\gamma^+$ (stars) and $\gamma^-$ (squares), the superscripts "$\pm$" corresponding to the circular polarizations $\sigma^+$ and $\sigma^-$. Lines are guides to the eye. Inset: temperature dependencies of $P_c = (\gamma_{ex}^+ - \gamma_{ex}^-)/(\gamma_{ex}^+ + \gamma_{ex}^-)$ (squares) and the Brillouin function $\tanh(\mu_B |g_e| B / 2 k_B T)$ for an electronic g-factor $g_e = -1.64$ (solid curve).

Figure 4 shows that at low temperatures $\gamma^+ > \gamma^-$, then $\gamma^+$ decreases with increasing temperature, while $\gamma^-$ increases. Such behavior of the homogeneous broadening parameters $\gamma^+$ and $\gamma^-$ with temperature could be ascribed to a spin-dependent exchange process in exciton-electron scattering. Such a scattering mechanism of this was considered in Ref.[8,9]. In the presence of magnetic fields when the 2DEG is completely spin polarized the exciton-electron scattering amplitude will depend strongly on the spin state of the exciton. For $\sigma^-$ polarization with total momentum $-1$ the electron spin of the exciton is parallel to the spin of the background electrons and we have no exchange contribution to the scattering, the initial and the final state is the same. But for the exciton in $\sigma^+$ polarization with a total momentum $+1$ due to the exchange of the electrons the electron in the final state will appear in the upper Zeeman sublevel and the exciton will be in the state with total momentum $+2$. Consequently, the exciton damping in $\sigma^+$ polarization should be higher than in $\sigma^-$ polarization.

To summarize, the homogeneous linewidth $\gamma^+$ for reflection of the $\sigma^+$ polarized component is found to be larger than the linewidth $\gamma^-$ of the $\sigma^-$ polarized component. Both parameters can be written in the form $\gamma^\pm = \gamma_{dir} + \gamma_{ex}^\pm$. Here, $\gamma_{ex}^+$ and $\gamma_{ex}^-$ describe contributions of spin-dependent exchange scattering that are believed to be proportional to the number of electrons with spins ($\uparrow$) or ($\downarrow$), respectively. The contribution $\gamma_{dir}$ is ascribed to spin-independent scattering.

At 1.6 K in a magnetic field electrons occupy the lowest Zeeman level with spin ($\uparrow$). In this case, in the $\sigma^-$ polarized component the contributions of exchange scattering are zero, and we find that the value of homogeneous broadening due to direct scattering is $\hbar\gamma_{dir} = \hbar\gamma^- = 0.15$ meV and that the total value of homogeneous broadening due to the exchange scattering is $\hbar\gamma_{ex} = \hbar\gamma_{ex}^+ + \hbar\gamma_{ex}^- = \hbar\gamma^+ - \hbar\gamma^- = 0.95$ meV.

At higher temperatures the upper Zeeman sublevel is occupied as well, and the number of electrons in the spin states ($\downarrow$) increases. The number of electrons with spins ($\uparrow$) and ($\downarrow$) become close to each other with increasing temperature and coincide above T=20 K, consequently, $\gamma^+ = \gamma^-$ in this case. The behavior of exchange scattering with temperature should reflect the polarization properties of quasi-2D electron gas. In the inset of Fig. 4 we compare the measured temperature dependence of $P_c = (\gamma_{ex}^+ - \gamma_{ex}^-)/(\gamma_{ex}^+ + \gamma_{ex}^-)$, being a measure of the quasi-2D electron gas polarization, and the Brillouin function $\tanh(\mu_B |g_e| B / 2 k_B T)$ calculated with the electronic g-factor $g_e = -1.64$ [24]. A good agreement between the experimental and theoretical data has been achieved without any fitting parameters.

Note that the suggested procedure of homogeneous and inhomogeneous linewidth separation is based on the effect of the homogeneous linewidth dependence on the circular polarization of the probing light in magnetic fields. The accuracy of the separation of the homogeneous and inhomogeneous contributions to the total linewidth is defined by the difference of the homogeneous linewidth in two circular polarizations. In undoped QW structures the exciton reflectivity line width has the same values in $\sigma^+$ and $\sigma^-$ polarizations and this method is not applicable.

## 5. Conclusion

We have elaborated a systematic procedure for analyzing inhomogeneous broadening of exciton resonant reflection spectra in quantum wells. Experimentally, the proposed method exploits the effect of circular polarization of probing light on the homogeneous broadening of the exciton linewidth, in contrast to inhomogeneous broadening, when a QW contains a 2DEG in an applied magnetic field. Theoretically, a phenomenological model has been formulated that ascribes the inhomogeneous broadening of the linewidth to fluctuations of the QW thickness followed by the fluctuations of the exciton transition energy through the carrier confinement. These ideas have been applied to analyze magneto-reflection spectra revealing inhomogeneous broadening of excitons in single quantum well CdTe/(Cd,Mg)Te heterostructures, containing a 2DEG with an electron density $8\times 10^{10}$ cm$^{-2}$ in magnetic fields of 5.5 T. The results presented here are typical and have been observed in other modulation doped samples. As a result, the values of excitonic parameters (resonance energy, radiative and nonradiative damping rates, and, especially, the parameter of inhomogeneous broadening) have been obtained by this method. The measured temperature dependencies of the nonradiative damping rate and the parameter of inhomogeneous broadening have been discussed as well. An estimation shows that the observed value of inhomogeneous broadening linewidth corresponds to a standard deviation of the SQW thickness on an atomic monolayer scale. Fluctuations of this order have been deduced for other high quality samples previously [27].

**Acknowledgments:** This work was supported in part by the Deutsche Forschungsgemeinschaft (grant No. Os98/6) and the Russian Foundation for Basic Research (grants Nos. 98-02-18219 and 00-02-16924). D.R.Y. acknowledges support of the Deutsche Forschungsgemeinschaft through SFB 410.

________________________________________________________________